\documentstyle[12pt,epsf]{article}
\newcommand{\dis}{\displaystyle}

\newcommand{\gsim}{\stackrel{>}{_\sim}}

\begin{document}
%\tighten

\thispagestyle{empty}
\begin{flushright}
IASSNS-HEP 97/41\\
SLAC-PUB-7448\\
{\tt hep-ph/9705253}\\
April 1997

\end{flushright}
\vspace*{1cm}
\centerline{\Large\bf Corrections of Order $\Lambda^2_{QCD}/m^2_c$}
\centerline{\Large\bf to Inclusive Rare B Decays
\footnote{
Work supported by the Department of Energy under contract
DE-AC03-76SF00515,  
NSF-KOSEF Bilateral Grant, KOSEF Purpose-Oriented 
Research Grant and SRC-Program, Ministry of Education Grant BSRI 97-2410,
the Monell Foundation and the Seoam Foundation Fellowships.}}
\vspace*{1.5cm}
\centerline{{\sc G. Buchalla${}^a$, G. Isidori${}^{a,b}$
%\footnote{On leave of absence from Laboratori Nazionali di Frascati, INFN, Italy.}
 \mbox{and} S.-J. Rey${}^{c,d}$} }
\bigskip
\centerline{\sl ${}^a$Stanford Linear Accelerator Center}
\centerline{\sl Stanford University, Stanford, CA 94309, U.S.A.}
\centerline{\sl ${}^b$INFN, Laboratori Nazionali di Frascati, 
                I-00044 Frascati, Italy}
\centerline{\sl ${}^c$School of Natural Sciences, 
              Institute for Advanced Study}
\centerline{\sl Olden Lane, Princeton, NJ 08540, U.S.A.}
\centerline{\sl ${}^d$Department of Physics, Seoul National University, 
Seoul 151-742 Korea}

\vspace*{1cm}
\centerline{\bf Abstract}
\vspace*{0.2cm}
\noindent 
We calculate nonperturbative ${\cal O}(\Lambda^2_{QCD}/m^2_c)$
corrections
to the dilepton invariant mass spectrum and the forward--backward
charge asymmetry in $B\to X_se^+e^-$ decay
using a heavy quark expansion approach. The method has recently been
used to estimate long--distance effects in $B\to X_s\gamma$. We 
generalize this analysis to the case of nonvanishing photon
invariant mass, $q^2\not= 0$, relevant for the rare decay mode
$B\to X_se^+e^-$. In the phenomenologically interesting $q^2$ region
away from the $c\bar c$ resonances, the heavy quark expansion approach
should provide a reasonable description of possible nonperturbative 
corrections. In particular this picture is preferable to the 
model--dependent approach relying on the tails of Breit--Wigner resonances, 
which 
has been employed so far in the literature to account for these effects.
We find that the ${\cal O}(\Lambda^2_{QCD}/m^2_c)$ corrections to the 
dilepton invariant mass spectrum and to the forward--backward asymmetry 
in $B\to X_se^+e^-$ amount to several percent at most for
$q^2/m^2_b\;\raisebox{-.4ex}{\rlap{$\sim$}} \raisebox{.4ex}{$<$}\; 0.3$
and
$q^2/m^2_b\;\raisebox{-.4ex}{\rlap{$\sim$}} \raisebox{.4ex}{$>$}\; 0.6$.
The ${\cal O}(\Lambda^2_{QCD}/m^2_c)$ correction to the $B\to X_s\gamma$
decay rate is also computed and found to be $+3\%$, which agrees in
magnitude with previous calculations.
Finally, we comment on long--distance effects in $B\to X_s\nu\bar\nu$,
which in this case are extremely suppressed due to the absence of
virtual photon contributions.
\vfill

\newpage
\pagenumbering{arabic}

\section{Introduction}

Inclusive rare decays of $B$ mesons, such as $B\to X_s\gamma$,
$B\to X_se^+e^-$ or $B\to X_s\nu\bar\nu$, provide important
opportunities to test the Standard Model of flavor physics.
These processes are particularly suited for this purpose, since
they occur only at the loop level and
are dominated by contributions that are reliably calculable in
perturbation theory. Heavy quark ($1/m_b$) expansion and 
renormalization group improved perturbative QCD form a solid
theoretical framework to describe these decays. The rates of
the $B$ meson decays are essentially determined by those of free
$b$ quarks, which have been computed at next--to--leading order (NLO)
in QCD for $B\to X_s\gamma$ \cite{GHW,CMM}, 
$B\to X_se^+e^-$ \cite{MIS,BM} and $B\to X_s\nu\bar\nu$ \cite{BB2}.
The first subleading ${\cal O}(1/m^2_b)$ power corrections in the
heavy quark expansion have also been studied and are in general well
under control \cite{FLS,AHHM,BBSUV,GLN}. 
However, to properly assess the prospects for precise tests of the
Standard Model and its extensions (see for instance \cite{HW}), 
further possible sources of theoretical
uncertainty, beyond those from $1/m_b$ expansion or QCD perturbation
theory, have to be investigated.
\\
It had been realized that there are, in principle, long--distance
contributions to $B\to X_s\gamma$ and $B\to X_se^+e^-$ related to the
$c\bar c$ intermediate state, which go beyond QCD perturbation theory
and which are also not described by nonperturbative higher order
contributions in the $1/m_b$ expansion. They originate from the
quark level process $b\to sc\bar c\to s\gamma$ and have previously been
estimated using model calculations based on $c\bar c$ resonance
exchange. For $B\to X_s\gamma$ one may have 
$b\to s\Psi\to s\gamma$, where the $\Psi$ (or a higher resonance)
converts into an on--shell photon ($q^2=0$). This requires
extrapolating the $\Psi$ couplings to far off--shell values
(from $q^2=M^2_\Psi$ to $q^2=0$). {}From estimates perfomed along these
lines effects of typically $\sim 10\%$ at the amplitude level were
found in \cite{DHT,SOA}.
Although the accuracy of this approach is difficult to quantify, a sizable
uncertainty from this source could not strictly be excluded.
\\
In the case of $B\to X_se^+e^-$ \footnote{For simplicity we write
$B\to X_se^+e^-$ throughout the paper with the understanding that our
discussion is equally applicable to $B\to X_s\mu^+\mu^-$.}, 
the intermediate $c\bar c$ resonances
can be on--shell for appropriate values $q^2$ of the dilepton invariant
mass. In this region of $q^2$, the resonance contributions form a very
large background to the flavor--changing neutral current signal.
Experimental cuts are therefore necessary to remove this part of the 
$q^2$--spectrum in the analysis of $B\to X_se^+e^-$, which is then
restricted to the range of $q^2$ below (and in principle also above)
the resonance region. Far from the resonances the situation with respect
to long--distance effects is analogous to the case of $B\to X_s\gamma$.
The tails of the resonances can extend into this region, however, the
Breit--Wigner form usually employed in modeling the resonance peaks
cannot be expected to give a correct description of the tail region
away from the peak.
Model--dependent treatments of the long--distance effects related to
the tails of $c\bar c$ resonances have been discussed in the literature
\cite{AHHM,AMM,AH}, but again cannot be considered as fully conclusive.
\\
Recently, it has been proposed to treat in a model--independent
way, employing a heavy quark expansion in inverse powers of the
charm--quark mass, the nonperturbative contributions in $B\to X_s\gamma$
related to the intermediate $c\bar c$ state \cite{VOL,KRSW,LRW,GMNP}.
This interesting approach, which leads to ${\cal O}(\Lambda^2_{QCD}/m^2_c)$ 
corrections, has the advantage of being well defined and systematic.
It avoids problems of double--counting of certain contributions, which
are present in model descriptions involving both charm--quark and hadronic
charm degrees of freedom simultaneously. Furthermore, for $c\bar c$ states
far off--shell, the quark picture appears to be more appropriate than
a description in terms of hadronic resonances ($\Psi$, $\Psi'$, ...).
Irrespective of the ultimate numerical accuracy that can be expected
from an expansion in $\Lambda_{QCD}/m_c$, we believe that this method can 
still yield a useful order of magnitude estimate of the effect, which is 
at least complementary, and probably superior, to alternative 
model--dependent calculations.
\\
The main purpose of the present article is to study nonperturbative
long--distance effects in $B\to X_se^+e^-$ in a model--independent way,
using the methods previously applied to $B\to X_s\gamma$. We will
calculate the ${\cal O}(\Lambda^2_{QCD}/m^2_c)$ corrections to the
dilepton invariant mass spectrum and to the forward--backward charge
asymmetry in $B\to X_se^+e^-$.
We also obtain the ${\cal O}(\Lambda^2_{QCD}/m^2_c)$ correction to the
branching ratio for $B\to X_s\gamma$ as a special case of this
analysis. Contrary to model estimates (involving $c\bar c$ resonances)
both the magnitude and the sign of the long--distance effect are
calculable and we find 
$\Delta B(B\to X_s\gamma)/B(B\to X_s\gamma)=+0.03$. Our result agrees
in magnitude with previous calculations \cite{VOL,LRW,GMNP}.
\\
We remark that in comparison to the case of $B\to X_s\gamma$ and 
$B\to X_se^+e^-$, long--distance effects in $B\to X_s\nu\bar\nu$ are
strongly suppressed by factors of ${\cal O}(m^2_c/M^2_Z)$. Obviously
the reason is that neutrinos do not couple to photons but
only to weak bosons, 
whose contributions are likewise completely negligible
for the long--distance effects in $B\to X_s\gamma$ and $B\to X_se^+e^-$.
{}From these observations it is clear that $B\to X_s \nu\bar\nu$ is
a particularly clean channel from a theoretical point of view.

The paper is organized as follows. After this Introduction, in section 2,
 we briefly review nonperturbative effects in $B\to X_s\gamma$
and $B\to X_se^+e^-$. We also
 discuss the $\bar sb$--photon--gluon vertex,
which will be of central importance for our subsequent calculations.
Using the results of section 2 in the limiting case of an on--shell photon,
we derive the ${\cal O}(\Lambda^2_{QCD}/m^2_c)$ correction to
the $B\to X_s\gamma$ rate in section 3. The main results of our paper,
the ${\cal O}(\Lambda^2_{QCD}/m^2_c)$ effects on the dilepton invariant
mass spectrum and the forward--backward asymmetry in $B\to X_se^+e^-$,
are calculated and discussed in section 4. Section 5 contains some
comments on long--distance effects in $B\to X_s\nu\bar\nu$. A short
summary is presented in section 6.

\section{Nonperturbative Effects in $B\to X_s\gamma^*$  and
the $\bar sb\gamma g$--Vertex}

\begin{figure}[t]
   \vspace{0cm}
   \epsfysize=6cm
   \epsfxsize=12cm
   \centerline{\epsffile{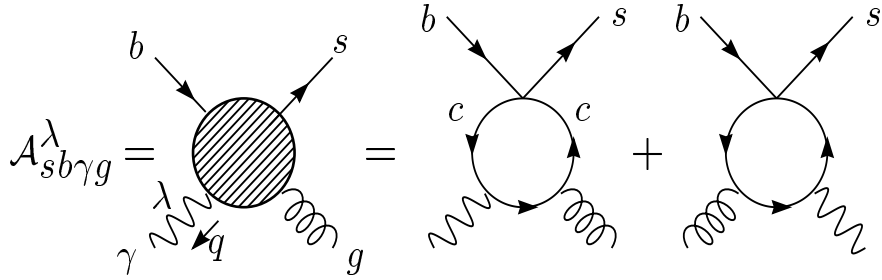}}
   \vspace*{-1.5cm}
\caption{\label{fig1} The $\bar sb\gamma g$ vertex at lowest
order in QCD.} 
\end{figure}

As mentioned before, $B\to X_s\gamma$ and $B\to X_se^+e^-$ are
dominated by perturbatively calculable short--distance contributions,
including those from virtual top quarks. At NLO in QCD the intrinsic
uncertainty of the perturbative calculation, indicated by residual
renormalization scale dependence, is about $\pm 6\%$ for $B\to X_s\gamma$
\cite{GHW,CMM} and of similar magnitude (or somewhat smaller) for the
$q^2$--spectrum in $B\to X_se^+e^-$ \cite{BM}. It is useful to keep
these numbers in mind for comparison with uncertainties from other
sources. 
Several effects that are beyond a perturbative treatment lead to
corrections to the purely partonic decay picture and have to be
considered for a more complete assessment of the total theoretical
uncertainty:
\begin{itemize}
\item
{\it Power corrections in $1/m_b$:}
Subleading corrections in the heavy quark expansion start at 
${\cal O}(\Lambda^2_{QCD}/m^2_b)$ and amount typically to several
percent. The effects are calculable, generally well under control, and
can be taken into account in a complete analysis \cite{FLS,AHHM}. 
An exception is
the photon energy spectrum in $B\to X_s\gamma$ \cite{NEU}
and the endpoint region of the dilepton
invariant mass ($q^2$) spectrum in $B\to X_s e^+e^-$ \cite{AHHM}, 
where the $1/m_b$ expansion breaks down. 
This is not a problem for the integrated total $B\to X_s\gamma$ rate 
and for the $B\to X_se^+e^-$ spectrum in the lower $q^2$ region.
It is more troublesome for the shape of the photon energy spectrum in
$B\to X_s\gamma$, which is needed to extract $B(B\to X_s\gamma)$ from
the data. At present this fact still introduces some amount of model
dependence in the measurement of this quantity \cite{AG}.
\item
{\it On--shell $c\bar c$ resonances:}
In the case of $B\to X_s\gamma$ there is a background from the process
$B\to X_s\Psi$, with the (on--shell) $\Psi$ subsequently decaying
into a photon plus light hadrons. However, the energy 
spectrum of photons from this cascade process peaks at 
lower values than the one from the short--distance decay $B\to X_s\gamma$. 
In the window $2.2$~GeV $< E_\gamma < 2.7$~GeV used to measure
$B(B\to X_s\gamma)$ \cite{CLEO} this background is fairly small and can be 
corrected for.
In the case of $B\to X_se^+e^-$ the $c\bar c$ resonance background
shows up as large peaks in the dilepton invariant mass spectrum. As
discussed in the Introduction, it has to be removed by appropriate cuts. 
\item
{\it Off--shell $c\bar c$ nonperturbative effects:}
A particular nonperturbative contribution to the short--distance process
$B\to X_s\gamma$ comes from the diagrams shown in Fig. \ref{fig1}.
It arises when the gluon is soft and interacts with the spectator cloud
of the $b$ quark inside the $B$ meson. This is the effect 
pointed out in \cite{VOL,KRSW} and further discussed
in \cite{LRW,GMNP}. The same mechanism affects also
$B\to X_se^+e^-$, in which case the photon invariant mass is
$q^2\not= 0$.
\end{itemize}

It is this last point we would like to focus on in the present paper.
Since the gluon is very soft and interacting with the light spectator
cloud, the process in Fig. \ref{fig1} corresponds to a $c\bar c$ pair
from $b$ decay converting into a (on-- or off--shell) photon.
Previous attempts to describe long--distance contributions based on
the conversion of intermediate $(c\bar c)$ resonances into a
photon (for both $B\to X_s\gamma$ and $B\to X_se^+e^-$) can be viewed
as model calculations aimed at addressing essentially the same effect.
\\
Staying with the quark picture, one may evaluate the diagrams in
Fig. \ref{fig1} directly. To first order in the gluon momentum
$k$ (with $k^2=0$), the vertex function can be written as 
\begin{equation}\label{asbgg} 
{\cal A}^\lambda_{sb\gamma g} = -i\frac{G_F}{\sqrt{2}}\lambda_c
C_2 e g Q_c \bar s\gamma_\mu(1-\gamma_5)G_{\alpha\varrho}b
\ \frac{F(r)}{24\pi^2 m^2_c}
\left[ \varepsilon^{\alpha\beta\mu\varrho}q_\beta q^\lambda-
\varepsilon^{\alpha\lambda\mu\varrho} q^2+
\varepsilon^{\beta\lambda\mu\varrho} q_\beta q^\alpha\right]~.
\end{equation}
Here $^\lambda$ is the Lorentz index of the photon, $q$ the photon
momentum and $G_{\alpha\varrho}=G^a_{\alpha\varrho}T^a$ the gluon
field strength. $C_2$ is the Wilson coefficient of the operator
$Q_2=(\bar sc)_{V-A}(\bar cb)_{V-A}$ in the standard effective weak
Hamiltonian. In the leading logarithmic approximation,
\begin{equation}\label{c2lla}
C_2=\frac{1}{2}
\left[\left(\frac{\alpha_s(M_W)}{\alpha_s(m_b)}\right)^{6/23}+
\left(\frac{\alpha_s(M_W)}{\alpha_s(m_b)}\right)^{-12/23}\right]~.
\end{equation}
Furthermore, $\lambda_c=V^*_{cs}V_{cb}$ and $Q_c=+2/3$ is the 
charm--quark electric charge. The form factor $F(r)$ as a function of
\begin{equation}\label{rdef}
r=\frac{q^2}{4 m^2_c}
\end{equation}
is given by
\begin{equation}\label{frl1}
F(r)=\frac{3}{2r}\left\{ \begin{array}{ll}
\dis\frac{1}{\sqrt{r(1-r)}}\arctan\sqrt{\frac{r}{1-r}} 
   -1 &  \qquad\qquad 0< r < 1~, \\
 \dis\frac{1}{2\sqrt{r(r-1)}}\left(
\ln\frac{1-\sqrt{1-1/r}}{1+\sqrt{1-1/r}}+i\pi\right)-1 &
\qquad\qquad r > 1~. \end{array} \right. 
\end{equation}
Our sign conventions are such that $\varepsilon^{0123}=+1$ and the
covariant derivative is
\begin{equation}\label{dcov}
D_\mu=\partial_\mu-i g T^a A^a_\mu + i e Q_f A_\mu~.
\end{equation}

The diagrams in Fig.~\ref{fig1} generate a finite amplitude.
However, in addition to the gauge--invariant $m_c$--dependent 
structure in (\ref{asbgg}), also terms proportional to 
$\varepsilon^{\alpha\lambda\mu\varrho}(q-k)_\alpha$ appear. 
We have not included these terms in ${\cal A}^\lambda_{sb\gamma g}$
because they are cancelled by the top--quark contribution
via the GIM mechanism. This can be easily seen by considering the
hierarchy $m_c\ll m_t\ll M_W$, which is sufficient since we are not
interested in the detailed form of the top--quark contribution to
${\cal A}^\lambda_{sb\gamma g}$ and neglect ${\cal O}(1/m^2_t)$ terms anyway.
We note also that by virtue of the GIM cancellation between top-- and
charm--quark contributions, the integrals are all manifestly convergent and
the full calculation, including the Dirac algebra, can be safely
performed in four dimensions.

Eq. (\ref{asbgg}) will be sufficient for our actual calculations
of ${\cal O}(\Lambda^2_{QCD}/m^2_c)$ effects in $B\to X_se^+e^-$ and
$B\to X_s\gamma$. It is, however, useful to also consider the more
general result for the $sb\gamma g$--vertex obtained without expanding
in $k\cdot q$. This will allow us to get some idea about the reliability
of the lowest order approximation and its stability against higher
order effects. 
Of course, there are further contributions, beyond those from higher
powers in $k\cdot q$. For instance, one could have terms with more than
one gluon field. But these are suppressed by additional powers of
$\Lambda^2_{QCD}/m^2_c$ and we shall neglect them. 
The vertex function (valid to all orders in $q^2$ and
$k\cdot q$, but with $k^2=0$) is then given by
\begin{eqnarray}
{\overline{\cal A}}_{sb\gamma g} &=& -i\frac{G_F}{\sqrt{2}}\lambda_c
C_2 e g Q_c \bar s\gamma_\mu(1-\gamma_5)T^ab
\ \frac{1}{48\pi^2 m^2_c} \varepsilon^{\alpha\lambda\mu\varrho} 
\nonumber \\ &&  \times \left[ {\overline F}(r+t,t) 
( G^a_{\lambda\sigma}\partial^{\sigma}F_{\varrho\alpha} +
 F_{\lambda\sigma}\partial^{\sigma}G^a_{\varrho\alpha} +
 G^a_{\alpha\varrho}\partial^{\sigma}F_{\sigma\lambda} )
 + {\overline F}(r,-t)
 G^a_{\alpha\varrho}\partial^{\sigma}F_{\sigma\lambda} \right]~,
\label{asbggt} 
\end{eqnarray}
where 
\begin{equation}\label{t_def}
t=\frac{k\cdot q}{2 m^2_c}
\end{equation}
and $F^{\alpha\beta}$ is the electromagnetic field strength. 
The form factor ${\overline F}$
is given by\footnote{~Note that the 
form factor ${\overline F}$ can be written as 
${\overline F}(r,t)=-384\pi^2 m^2_c {\widetilde C_{20}}(q^2,k\cdot q)$
in terms of the subtracted three--point function ${\widetilde C_{20}}$ 
defined in \cite{DEIN}; the explicit expression of ${\widetilde C_{20}}$ can be
found in the second ref. in \cite{DEIN}.} 
\begin{equation}
{\overline F}(r+t,t) =6\int^1_0dz\left\{-\frac{1-z}{t}-
\frac{1-4 z(1-z)r}{4zt^2}\ln\frac{1-4z(1-z)(r+t)}{1-4z(1-z)r}\right\}~.  
\label{Fbardef}
\end{equation}
The following relations hold
\begin{equation}
\label{FF1}
{\overline F}(r,0)=F(r)~, \qquad\qquad\qquad F(0)=1~.
\end{equation}

\noindent
Expanding ${\overline F}$ in powers of $t$,
%\begin{equation}
%F_1(t,r) = F(r) + \sum_{n=1}^\infty t^n c_{1,n}(r) \nonumber \qquad 
%F_2(t,r) = F(r) + \sum_{n=1}^\infty t^n c_{2,n}(r) \label{taylor}~,
%\end{equation}
(\ref{asbggt}) can be understood 
as a sum over an infinite
number of operators involving all powers of the gluon momentum
\cite{LRW,GMNP}. 
Unfortunately, for the physical processes we want to study, only the 
lowest order contribution can be calculated.
The higher derivative operators lead to matrix elements that are
unknown and one has to resort to a merely qualitative discussion
in this case.

Since $q$ is of order $m_b$
and we are interested in the limit where $k\sim {\cal O}(\Lambda_{QCD})$,
we expect the matrix elements of these operators to be suppressed 
by corresponding powers of $m_b \Lambda_{QCD} /m_c^2$ 
(suppression of $t^{n+1}$ with respect to $t^n$) and 
$\Lambda_{QCD}/m_b$ (suppression of $F\partial G$ versus 
$G\partial F$). 
If both of the following inequalities were satisfied 
\begin{eqnarray}
m_b \gg \Lambda_{QCD}~, \label{approx1}\\
m_c^2 /m_b \gg \Lambda_{QCD}~, \label{approx2}
\end{eqnarray}
we could neglect all the operators involving higher powers of the 
gluon momentum and use (\ref{asbgg}) instead of (\ref{asbggt}). 
However, whereas (\ref{approx1}) is well justified, (\ref{approx2}) is 
not a priori a good approximation in the real world. 
Thus we can safely neglect 
the $F\partial G$ term in (\ref{asbggt}) but we must have a closer look 
at the ${\cal O}(t^n)$ corrections to (\ref{asbgg}).

\noindent
In the $B\to X_s\gamma$ case ($q^2=0$) these corrections are estimated 
to be small \cite{LRW,GMNP}. Indeed, even though $k\cdot q/m_c^2 \sim
{\cal O}(1)$, the Taylor expansion of ${\overline F}(t,t)$ in terms of 
$k\cdot q/m_c^2$ involves small coefficients
\begin{equation}
{\overline F}(t,t)= 1+\frac{4}{15}\frac{k\cdot q}{m^2_c}+
\frac{3}{35}\left(\frac{k\cdot q}{m^2_c}\right)^2+
\frac{16}{525}\left(\frac{k\cdot q}{m^2_c}\right)^3+
\frac{8}{693}\left(\frac{k\cdot q}{m^2_c}\right)^4+\ldots
\end{equation}
A typical value for $t$ would be $t\approx 0.3$ \cite{LRW}. Then we have
${\overline F}(0.3,0.3)\approx 1.2$, compared with ${\overline F}(0,0)=1$.
This indicates that the summed contribution of the $t^n$--terms 
($n\geq 1$) is probably not too important, owing to the small numerical
coefficients in the series \cite{LRW,GMNP}. Of course, this is a
rather crude argument, since the matrix elements of the higher derivative
operators may be weighted differently for different $n$. Furthermore,
coherent addition of all the terms may not be correct either, 
but is at least a conservative assumption.
(Indeed, the matrix element of the $n=1$ operator turns out to be
vanishing in the case of $B\to X_s\gamma$ \cite{GMNP}.) Still
this estimate, despite being rough, provides a certain measure of the
possible importance of neglected effects, exploiting the information
contained in the form factor.

In the same spirit one may investigate the general case $q^2\not= 0$.
The ratio ${\overline F}(r+t,t)/{\overline F}(r,0)$, measuring the
sensitivity to a nonvanishing $t$, reads
$\approx 1.3$, $1.4$, $1.5$, $1.75$ for $r=0.3$, $0.4$, $0.5$ and
$0.6$, respectively, using $t=0.3$
(these values of $r$ correspond to $q^2/m^2_b\simeq 0.1$, $0.14$,
$0.17$ and $0.2$).
This suggests that the expansion in $t$ makes sense
provided that $q^2$ is
far enough from the $4m_c^2$ threshold.
On the other hand, due to the physical 
cuts at $q^2=4m_c^2$ and $(k+q)^2=4m_c^2$
of the diagrams in Fig.~\ref{fig1}, the functions ${\overline F}$
diverge
near the $4m_c^2$ threshold. This breakdown of the gluon--momentum
expansion can be viewed as an indication of the appearance of 
large genuine long--distance effects due to the 
nearby $c\bar{c}$ resonances.   
Hence, we conclude that if
\begin{equation}
4m_c^2 -q^2 \gsim m_c^2
\end{equation}
the lowest order ($t=0$) result is still reasonable, as a sensible
approximation for smaller $q^2$ and as an order--of--magnitude estimate
closer to this bound.

So far we have been considering the region of $q^2$ below the
resonance domain. Above the resonances $r$ is parametrically large,
$r\sim q^2/m^2_c\approx m^2_b/m^2_c\gg 1$, and one may expand
${\overline F}(r+t,t)$ not only in $t$ but subsequently also in
inverse powers of $r$. Keeping only the leading term in $1/r$ in each
of the coefficients of $t^n$ one has
\begin{equation}\label{fstr}
{\overline F}(r+t,t)\approx\sum^\infty_{n=0}(-1)^{n+1}
\frac{3}{(n+2) r^{n+1}}t^n=-\frac{3}{2r}+\frac{t}{r^2}+\ldots
\end{equation}
This shows that the leading corrections to the $t=0$ result behave as a
series in powers of $t/r\sim k\cdot q/q^2\sim\Lambda_{QCD}/m_b$.
Numerically, using the full expression for ${\overline F}(r+t,t)$ 
and taking $r=2$ as a relevant example, we find that
${\overline F}(2-0.3,-0.3)=-1.40+1.10 i$. To be more conservative we have
here chosen $t<0$ so that all contributions add coherently. The result
is reasonably close to the $t=0$ value ${\overline F}(2,0)=-1.22+0.83 i$. 
Thus it appears justified to neglect the higher dimensional operators 
in this case and we expect the approximation $t=0$ to work 
in the large $q^2$ region of the spectrum.
\\
In the previous discussion we have restricted our attention to the
function ${\overline F}(r+t,t)$, but similar conclusions hold also
for ${\overline F}(r,-t)$. The latter contributes only for $q^2\not= 0$
and its expansion in $t$ shows a better behavior than the one of
${\overline F}(r+t,t)$.

At this point it is natural to address the question of 
the effect generated by the diagrams in Fig.~\ref{fig1} if
the charm quark 
inside the loop is replaced by an up quark. At first sight,
the explicit $m_c^2$ dependence of (\ref{asbggt}) would suggest that the
effect diverges as $1/m_q^2$ when $m_q\to 0$. This is still not a problem
for $B\to X_s\gamma^*$, where the up--quark contribution is suppressed by 
the CKM  hierarchy, but could be a dramatic effect in 
$B\to X_d\gamma^*$. However, it should be noted that
the $1/m_q^2$ behavior of (\ref{asbggt})
is correct only if the expansion of ${\overline F}$ in terms of 
$k\cdot q/m_q^2$ is allowed, which is not true for the up--quark
contribution. In this case, and considering $q^2=0$ for the moment, 
it is more appropriate to expand
${\overline F}(t,t)$ in inverse powers of $t$: 
the leading term in this expansion goes like $t^{-1}\sim m^2_q/k\cdot q$ 
and cancels the artificial 
$1/m_q^2$ divergence of (\ref{asbggt}). The operators generated 
by this expansion are nonlocal and their matrix elements 
cannot be estimated reliably. However, from naive dimensional counting, 
we expect the leading contribution to be of order $\Lambda_{QCD}/m_b$.
\\
For the up--quark contribution in $B\to X_{s,d}e^+e^-$ the situation
is, in a sense, even more favorable. If 
$q^2\;\raisebox{-.4ex}{\rlap{$\sim$}} \raisebox{.4ex}{$>$}\;(2\mbox{GeV})^2$,
one is above the region of $u\bar u$ resonances and $q^2$ 
itself provides the relevant short--distance scale. 
The leading long--distance contribution is then of the order
$\Lambda^2_{QCD}/q^2$. Corrections to this behavior arise as powers
of $t/r\sim k\cdot q/q^2\sim \Lambda_{QCD}/\sqrt{q^2}$, corresponding to
a series of local, higher--dimensional operators. These results follow
from (\ref{fstr}). The situation is analogous to the one for the 
high--$q^2$ end of the spectrum in the case of intermediate $c\bar c$,
except that for the $u\bar u$ case the operator product expansion
approach remains valid down to smaller $q^2$ due to the absence
of heavy resonances.

We finally remark that the diagrams in Fig.~\ref{fig1} actually 
contribute in two distinct ways to the decays $B\to X_s\gamma^*$.
First, the gluon can be soft and couple to the light cloud in the $B$
meson. This is the effect we are mainly interested in here and which we 
shall calculate, far from the resonance region, using 
(\ref{asbgg}). Second, the gluon, now not
necessarily soft, may be radiated from the charm--quark
 loop to end up in the
final state $X_s$. This process is calculable in perturbation theory,
it is infrared finite and contributes as part of the matrix element
calculation at NLO in $B\to X_s\gamma$ \cite{GHW,AGbrem}. 
For $B\to X_se^+e^-$ the situation
is analogous, except that the diagrams of Fig. \ref{fig1} would enter
only beyond the next--to--leading order 
(at `next--to--next--to--leading order').
The first of these processes is not contained within the second
one, which is entirely perturbative. Therefore in calculating
the nonperturbative correction there is no double counting of 
contributions already taken into account in the higher order perturbative
result. 

\section{$B\to X_s\gamma$}

In this section we re--derive the ${\cal O}(\Lambda^2_{QCD}/m^2_b)$
correction to $\Gamma(B\to X_s\gamma)$ using
the results of the previous chapter in the limit of an on--shell photon.
\\
By means of the optical theorem the total rate of a $B$ meson decay can be
expressed as the expectation value
\begin{equation}\label{gamt}
\Gamma=\frac{1}{2M_B}\langle B|{\cal T}|B\rangle
\end{equation}
of a transition operator ${\cal T}$ defined by
\begin{equation}\label{tdef}
{\cal T}=\mbox{Im}\ i\int d^4x\ T {\cal H}_{eff}(x){\cal H}_{eff}(0)~.
\end{equation}
Here ${\cal H}_{eff}$ is the low energy effective Hamiltonian
governing the decay under consideration. The states $|B\rangle$ in
(\ref{gamt}) are to be taken in conventional relativistic normalization
($\langle B|B\rangle=2EV$).
\\
Since we are analyzing a small correction to the full decay width, it
is sufficient to work to leading logarithmic accuracy and to neglect
consistently all relative ${\cal O}(\alpha_s)$ effects. In this
approximation we may write
\begin{equation}\label{heff01}
{\cal H}_{eff}={\cal H}^{(0)}_{eff}+{\cal H}^{(1)}_{eff}~,
\end{equation}
where
\begin{equation}\label{heff0}
{\cal H}^{(0)}_{eff}=-\frac{G_F}{\sqrt{2}}V^*_{ts}V_{tb}\ C_7
{\cal O}_7+ {\rm h.c.}~,
\qquad 
{\cal O}_7=
\frac{e}{8\pi^2}m_b \bar s\sigma^{\mu\nu}(1+\gamma_5)b\ F_{\mu\nu}~,
\end{equation}
represents the well known leading--log effective Hamiltonian for
$b\to s\gamma$ decay \cite{CFRS} (for a review see \cite{BBL}).
$C_7\simeq -0.30$ is the scheme independent (`effective') Wilson
coefficient, whose analytic form can be found for instance in \cite{BBL}.
\\
The correction ${\cal H}^{(1)}_{eff}$ can be derived from the
amplitude (\ref{asbgg}) in the limit where the photon is put on--shell. 
In this case the vertex (\ref{asbgg}) is equivalent to a local operator.
Performing the usual matching procedure one obtains
\begin{equation}\label{heff1}
{\cal H}^{(1)}_{eff}=\frac{G_F}{\sqrt{2}}V^*_{cs}V_{cb}\ C_2\ 
{\cal O}_{11}+ {\rm h.c.}~,
\end{equation}
\begin{equation}\label{q11}
{\cal O}_{11}=\frac{e Q_c}{48\pi^2 m^2_c}\bar s\gamma_\mu(1-\gamma_5)
g G_{\nu\lambda}b\ \varepsilon^{\mu\nu\varrho\sigma}\partial^\lambda 
F_{\varrho\sigma}~.
\end{equation}
To leading order only ${\cal H}^{(0)}_{eff}$ contributes to (\ref{gamt}),
(\ref{tdef}). One finds the well--known result 
($V^*_{cs}V_{cb}\simeq -V^*_{ts}V_{tb}$)
\begin{equation}\label{gbsg0}
\Gamma(B\to X_s\gamma)=\frac{G^2_F m^5_b}{192\pi^3}
(V^*_{cs}V_{cb})^2\ \frac{6\alpha}{\pi} C^2_7~.
\end{equation}
The interference between ${\cal H}^{(0)}_{eff}$ and
${\cal H}^{(1)}_{eff}$ leads to a correction  
of the transition operator
\begin{equation}\label{delt}
\Delta{\cal T}=-\frac{G^2_F m^5_b}{192\pi^3}(V^*_{cs}V_{cb})^2\
\frac{\alpha}{9\pi}\ \frac{C_2\ C_7}{m^2_c}\ \bar b g\sigma\cdot G b~,
\qquad \sigma_{\mu\nu}\equiv\frac{i}{2}[\gamma_\mu,\gamma_\nu]~.
\end{equation}
Using \cite{BBSUV} 
\begin{equation}\label{sigg}
\frac{\langle B|\bar b g\sigma\cdot G b|B\rangle}{2M_B}=
\frac{3}{2}(M^2_{B^*}-M^2_B),
\end{equation}
we finally obtain 
\begin{equation}\label{finres}
\frac{\Delta\Gamma(B\to X_s\gamma)}{\Gamma(B\to X_s\gamma)}=
-\frac{C_2}{36 C_7}\frac{M^2_{B^*}-M^2_B}{m^2_c}\approx +0.03~.
\end{equation}
This result suggests that the potentially
problematic nonperturbative effects in $B\to X_s\gamma$ should indeed be
negligible, re--enforcing the role of the $B\to X_s\gamma$ process
as a significant test of the Standard Model.
\\
We stress that the correction (\ref{finres}) has a definite sign,
contrary to previous estimates of nonperturbative 
$c \bar c$ effects based on off--shell 
resonance exchange \cite{DHT,SOA}. In this context
we note that for a positive matrix element (\ref{sigg}) and negative
$C_7$, the consistent covariant derivative, specifying the sign of
strong and electromagnetic couplings and thus of (\ref{finres}),
is the one given in (\ref{dcov}). 
\\
The nonperturbative correction in (\ref{finres}) may be compared with
model calculations of the long--distance effect.
Assuming that the dominant 
contribution to (\ref{finres}) is generated by 
the $\Psi$ exchange, the above result can be used to fix
sign and magnitude of the nonfactorizable $b s \Psi$ coupling 
at $q^2=0$ ($g_2$ in the notation of \cite{SOA}). The result thus 
obtained is consistent with the lower values given in \cite{SOA}
($g_2\sim 10^{-2}$).

\section{$B\to X_se^+e^-$}

In the present section we proceed to compute the nonperturbative
correction to the rare decay $B\to X_se^+e^-$.
The effective Hamiltonian for $b\to se^+e^-$ at next--to--leading order
has been reviewed in \cite{BBL}. In the following we will make use of
the results collected in this article and adopt its notation. At 
next--to--leading order, the amplitude for $B\to X_se^+e^-$ can be
written as
\begin{equation}\label{abse}
{\cal A}=-i\frac{G_F}{\sqrt{2}}\lambda_c\frac{\alpha}{2\pi}\left[
2 m_b C_7 J^{\kappa\nu}_7\frac{i q_\kappa}{q^2}\ \bar u\gamma_\nu v+
\tilde C^{eff}_9 J^\nu_9 \ \bar u\gamma_\nu v+
\tilde C_{10} J^\nu_9 \ \bar u\gamma_\nu\gamma_5 v\right]~,
\end{equation}
where
\begin{equation}\label{j7kn}
J^{\kappa\nu}_7=\langle X_s|\bar s\sigma^{\kappa\nu}(1+\gamma_5)b|B\rangle
\end{equation}
\begin{equation}\label{j9n}
J^{\nu}_9=\langle X_s|\bar s\gamma^\nu(1-\gamma_5)b|B\rangle
\end{equation}
and $u$, $v$ denote the electron and positron spinor, respectively.
$C_7$ and $\tilde C_{10}$ are Wilson coefficients; $\tilde C^{eff}_9$
is not strictly speaking a Wilson coefficient, since it includes also
contributions from the $b\to se^+e^-$ matrix elements of the
operators $Q_1$, ... $Q_6$ in the Hamiltonian \cite{BBL}.
However,
the amplitude expressed
in terms of $\tilde C^{eff}_9$ provides a convenient notation.
In particular, $\tilde C^{eff}_9$ already contains contributions from
intermediate charm--quark states entering the one--loop matrix elements
(these are similar to the graphs in Fig. \ref{fig1} but without the gluon
and a lepton line connected to the photon propagator).
\\
In writing (\ref{abse}) we have used 
$V^*_{ts}V_{tb}\approx -V^*_{cs}V_{cb}=-\lambda_c$. Note also that, to
calculate $B\to X_se^+e^-$ at NLO, only the leading logarithmic
approximation is necessary for $C_7$ ($\equiv C^{(0)eff}_{7\gamma}$
in the notation of \cite{BBL}).

The nonperturbative correction we want to evaluate stems from
the vertex function in (\ref{asbgg}), which yields the following
correction to the leading amplitude in (\ref{abse})
\begin{equation}\label{acbse}
{\cal A}^c=-i\frac{G_F}{\sqrt{2}}\lambda_c\frac{\alpha}{2\pi}
\frac{C_2 Q_c}{3 m^2_c}\frac{F(r)}{q^2}\ J_{G, \mu\alpha\varrho}\left[
\varepsilon^{\beta\lambda\mu\varrho}q_\beta q^\alpha -
\varepsilon^{\alpha\lambda\mu\varrho} q^2\right]\ \bar u\gamma_\lambda v
\end{equation}
with
\begin{equation}\label{jgmar}
J_{G,\mu\alpha\varrho}=\langle X_s|\bar s\gamma_\mu(1-\gamma_5)
g G_{\alpha\varrho}b|B\rangle~.
\end{equation}
Note that the term proportional to $q^\lambda$ in (\ref{asbgg})
vanishes when multiplied by $\bar u\gamma_\lambda v$ due to current
conservation.
\\
Adding ${\cal A}+{\cal A}^c$, squaring the amplitude, summing over
inclusive states $X_s$ and performing the necessary phase space
integration, one may calculate the differential decay rate for
$B\to X_se^+e^-$.
Defining $s=q^2/m^2_b$ and
\begin{equation}\label{rsdef}
R(s)=\frac{\frac{d}{ds}\Gamma(B\to X_se^+e^-)}{\Gamma(B\to X_ce\nu)}~,
\end{equation}
one has at NLO in QCD perturbation theory ($z=m_c/m_b$)
\begin{equation}\label{rsnlo}
R(s)=\frac{\alpha^2}{4\pi^2}\frac{(1-s)^2}{f(z)\kappa(z)}\left[
(1+2s)\left(|\tilde C^{eff}_9|^2+|\tilde C_{10}|^2\right)+4
\left(1+\frac{2}{s}\right) |C_7|^2 + 12 C_7 \mbox{Re} \tilde C^{eff}_9
\right]~.
\end{equation}
Here $f(z)$ is the phase space factor and $\kappa(z)$ the QCD correction
factor entering $\Gamma(B\to X_ce\nu)$; they can be found in \cite{BBL}.
\\
For the ${\cal O}(\Lambda^2_{QCD}/m^2_c)$ correction term, arising
from the interference of ${\cal A}^c$ with ${\cal A}$, we obtain
\begin{equation}\label{delrs}
\Delta R(s)=-\frac{\alpha^2}{4\pi^2}C_2
\frac{2(M^2_{B^*}-M^2_B)}{9 m^2_c}
\frac{(1-s)^2}{f(z)\kappa(z)}\ \mbox{Re} \left\{ F(r)\left[
C^*_7\frac{1+6s-s^2}{s}+\tilde C^{eff*}_9 (2+s)\right] \right\}~.
\end{equation}
For the evaluation of $\Delta R$ we have used the identity \cite{FN}
\begin{equation}\label{ident}
\langle B|\bar b\Gamma G_{\alpha\beta}b|B\rangle=\frac{1}{48}
\langle B|\bar b\sigma\cdot G b|B\rangle \ \ \mbox{tr}\
\{\Gamma(1+\not\! v)\sigma_{\alpha\beta}(1+\not\! v)\}~,
\end{equation}
which is valid to leading order in the $1/m_b$ expansion. 
Here $v$ denotes the $B$ meson four--velocity and $\Gamma$ an arbitrary
string of $\gamma$--matrices.
For handling the Dirac algebra throughout our calculations we have
used the program Tracer \cite{JL}.

\begin{figure}[t]
   \vspace{0cm}
   \epsfysize=10cm
   \epsfxsize=10cm
   \centerline{\epsffile{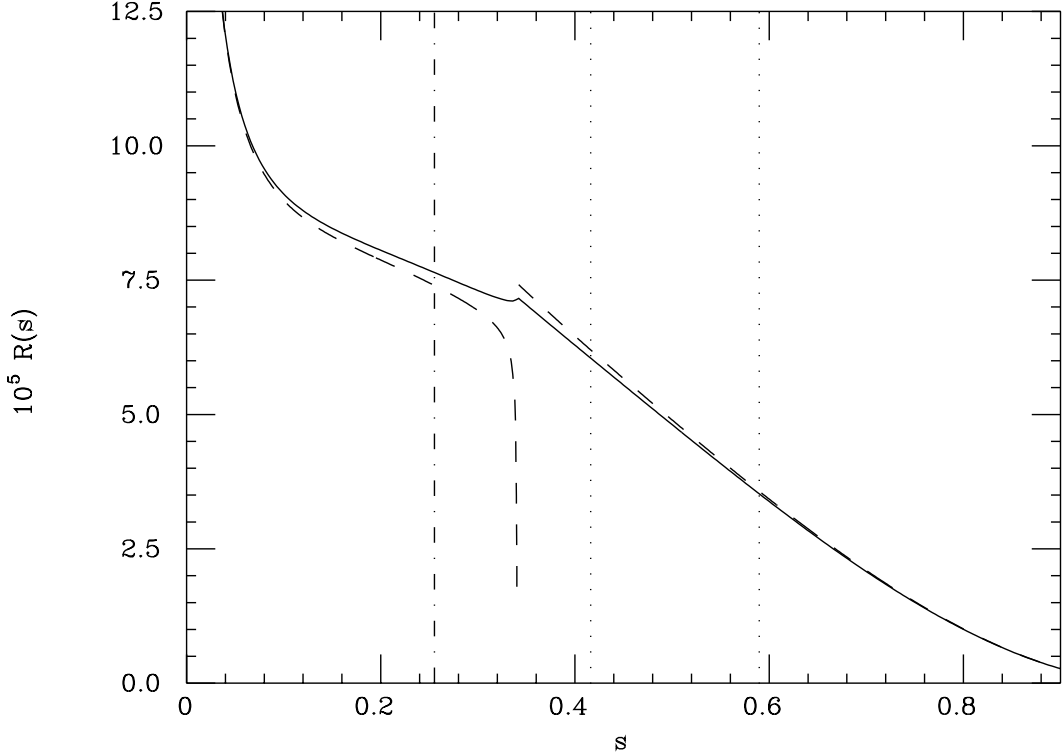}}
   \vspace*{-0.5cm}
\caption{\label{fig2} The dilepton invariant mass spectrum of 
$B\to X_se^+e^-$ normalized to the semileptonic width 
(\protect\ref{rsdef}). The solid curve is the short--distance result 
(\protect\ref{rsnlo}) whereas  
the dashed one includes the ${\cal O}(\Lambda^2_{QCD}/m^2_c)$ corrections.
The dotted lines indicate the position of the $\Psi$ 
and $\Psi'$ resonances. A dash--dotted line is shown 
at $q^2=3 m^2_c$. The corrections diverge
at $q^2=4 m^2_c$. In this plot we have used $m_c=1.4$ GeV
and $m_b=4.8$ GeV.} 
\end{figure} 

The final result is displayed in Fig.~\ref{fig2}, showing the corrected
and uncorrected $R(s)$. 
The conditions for the validity of the corrections have already been 
discussed in section 2. They are best satisfied for the high--$q^2$ end
of the spectrum, above the resonances, and for very low $q^2$. The
reliability of the approximation used deteriorates somewhat as $q^2$ is
increased from $\approx 0$ towards the resonance region. As an order
of magnitude estimate the calculation should still make sense around
$q^2\approx 3m^2_c$, indicated in Fig.~\ref{fig2} by the dash--dotted
line. For larger $q^2$ one gets too close to the lowest resonance and the
heavy quark expansion breaks down. Further up, at $q^2=4m^2_c$, the
correction develops a (unphysical) square root divergence (in our choice
$m_c=1.4$ GeV). The difference between $M_\Psi$ and $2 m_c$ is of
${\cal O}(\Lambda_{QCD})$ and corresponds to nonperturbative effects
that are beyond the control of our approximation.
\\
As can be noticed, the 
${\cal O}(\Lambda^2_{QCD}/m^2_c)$ corrections are very small 
in the region where this calculation can be trusted, that is below 
$3 m_c^2$ and above the resonance peaks. Actually, in the high $q^2$
region the effect is further suppressed since 
$F(r)$ decreases as $1/r$ for large $r$, as discussed in section 2. 
For $q^2 \sim {\cal O}(m_b^2)$ the corrections behave in fact more as
${\cal O}(\Lambda_{QCD}^2/m_b^2)$ instead of 
${\cal O}(\Lambda_{QCD}^2/m_c^2)$.
We also observe that for $s$ below $0.1$ the corrections become
vanishingly small due to a cancellation among the terms in (\ref{delrs})
($C_7$ is negative, $\mbox{Re} \tilde C^{eff}_9$ positive).
This is related to the fact that the correction $\Delta R(s)$ is negative
for $s$ above $0.1$, where $\tilde C^{eff}_9$ dominates, but positive
for $s$ close to zero, which essentially corresponds to the case of
$B\to X_s\gamma$. In fact, one may obtain (\ref{finres}) as a special
case of $\Delta R/R$ from (\ref{rsnlo}) and (\ref{delrs}) in the limit
$q^2\to 0$. 
\\
The sign of the effect for $s$ around $0.2$ is 
different from the one obtained by the approaches based on
resonance exchange \cite{AHHM,AMM,AH}. This is not a problem 
since the sign of the correction determined in model
calculations is not reliable far from the resonances. Moreover,
as discussed in the previous section, the corrections we are 
considering are essentially related to the nonfactorizable contributions 
of charmed resonances. The latter are usually considered in
$B\to X_s \gamma$, where the factorizable terms vanish,
but are always neglected in $B\to  X_s e^+e^-$. 
\\
The factorizable resonance contributions can be identified,
to the lowest order in $\alpha_s$, with diagrams as those in 
Fig.~\ref{fig1} but without gluons. The effect of such diagrams 
is positive and is already included in the NLO calculation
as the matrix element contribution to $\tilde C^{eff}_9$. As pointed 
out in \cite{LW} within  the context of 
exclusive decays, adding the factorizable resonance 
effects to the NLO calculation leads to a
double counting problem related to the simultaneous use of quark and
hadronic ($c\bar c$ resonances) degrees of freedom.
On the other hand, this is not the case 
for the $\Lambda_{QCD}^2/m_c^2$ effect we have evaluated, where we
have consistently used a pure quark level description.
Of course, some model--dependent treatment may still be useful
close to the $\Psi$--resonance (an interesting approach has been
presented in \cite{KS}). 
If one should attempt this, the
${\cal O}(\Lambda^2_{QCD}/m^2_c)$ effect could be used to fix the
small $q^2$ limit.

Another interesting quantity that can be measured in 
$B\to X_se^+e^-$ decays is the forward--backward charge 
asymmetry \cite{AMM}.  Normalizing $\Gamma(B\to X_se^+e^-)$ to the 
semileptonic width, as in (\ref{rsdef}), we define 
\begin{equation}\label{asdef}
A(s)=\frac{1}{\Gamma(B\to X_ce\nu)}
  \int_{-1}^1 d\cos\theta ~
 \frac{d^2 \Gamma(B\to X_se^+e^-)}{d s~ d\cos\theta}
\mbox{sgn}(\cos\theta)~,
\end{equation}
where $\theta$ is the angle between the $e^+$ and $B$ momenta 
in the dilepton center--of--mass frame. Proceeding similarly to 
the case of $R(s)$, the NLO perturbative result turns out to be
\begin{equation}\label{asnlo}
A(s)= - \frac{3\alpha^2}{4\pi^2}\frac{ (1-s)^2}{f(z)\kappa(z)} 
\mbox{Re} \left\{\tilde C^*_{10}
\left[ 2\ C_7 + s\ \tilde C^{eff}_9 \right] \right\}~,
\end{equation}
whereas the ${\cal O}(\Lambda^2_{QCD}/m^2_c)$ correction term, arising
from the interference of ${\cal A}^c$ and ${\cal A}$, is given by
\begin{equation}\label{delas}
\Delta A(s)= \frac{3\alpha^2}{4\pi^2}C_2
\frac{M^2_{B^*}-M^2_B}{36 m^2_c}
\frac{(1-s)^2}{f(z)\kappa(z)}\ \mbox{Re} \left\{\tilde C^*_{10}F(r) \right\}
\left( 1+3s \right)~.
\end{equation}
\begin{figure}[t]
   \vspace{0cm}
   \epsfysize=10cm
   \epsfxsize=10cm
   \centerline{\epsffile{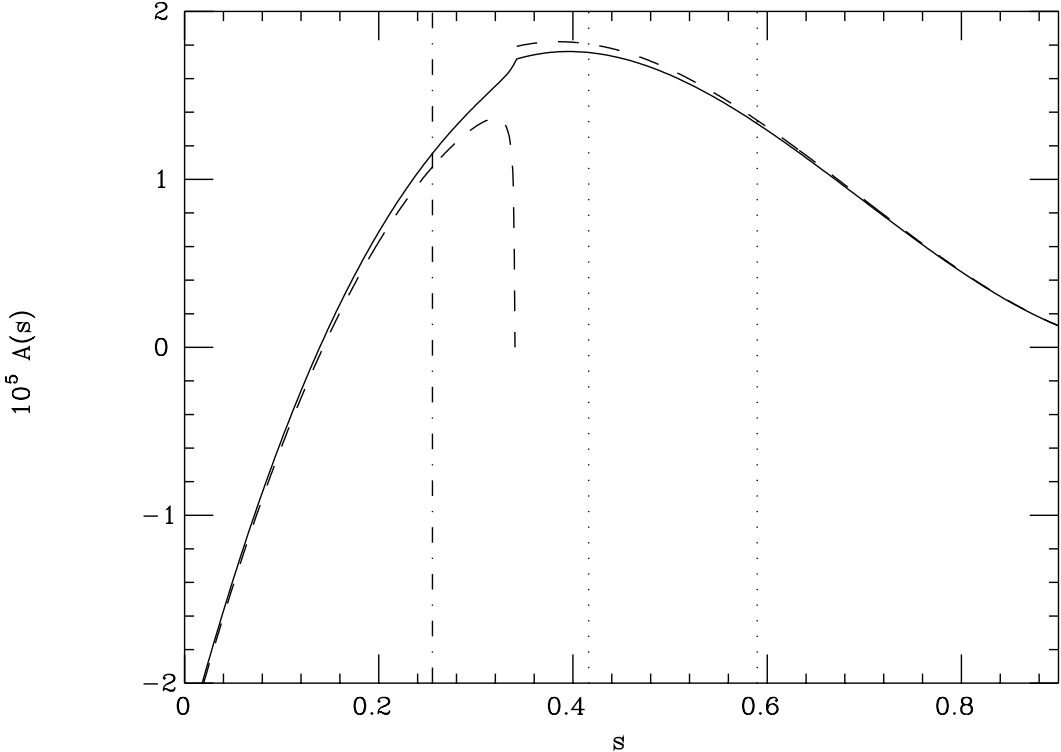}}
   \vspace*{-0.5cm}
\caption{\label{fig3} The differential forward--backward 
asymmetry defined in (\protect\ref{asdef}) 
as a function of the dilepton invariant mass. Notations are as in 
Fig.~\protect\ref{fig2}.} 
\end{figure} 
The corrected and uncorrected results for $A(s)$ 
are shown in Fig.~\ref{fig3}.
Concerning range of validity and size of the corrections similar
comments apply to $A(s)$ as to the spectrum $R(s)$ discussed above.
In addition we note that the small corrections to $A(s)$ and $R(s)$ have
a tendency to further cancel in the normalized asymmetry
$A(s)/R(s)$, a useful observable for direct comparison with experiment.

\section{$B\to X_s \nu \bar{\nu}$}

In $B\to X_s\nu\bar\nu$ there are no contributions from virtual
photons. Consequently the pattern of GIM cancellation is not
logarithmic (`soft'), as in $B\to X_se^+e^-$, but powerlike (`hard')
due to the exchange of heavy gauge bosons. Therefore it is clear that
long--distance effects in $B\to X_s\nu\bar\nu$ from charm quarks 
are additionally suppressed by a factor of order $m^2_c/M^2_W$.
In principle, however, also in this case long--distance contributions
related to the $c\bar c$ intermediate state do exist.
The largest of these is associated with the cascade process
$B \to X_s \Psi \to X_s \nu \bar{\nu} $.
In the following we shall try to estimate this 
effect to show more quantitatively 
to which extent the short--distance contribution dominates
in $B\to X_s \nu \bar{\nu}$.
\\
Using the relation $\langle 0 | \bar c \gamma^\mu c | 
\Psi \rangle = m_\Psi f_\Psi \epsilon^\mu$ and the coupling of the
$\bar c \gamma^\mu c$ current with the electroweak fields  
\begin{equation}
{\cal L}  = - e{\bar c}\gamma^\mu c \left[~{2\over 3} A_\mu
+{1\over4 \sin \theta_w \cos \theta_w }
\left(1-{8\over 3}\sin^2 \theta_w \right) Z_\mu \right]~,
\end{equation}
we obtain
\begin{equation}
R_{\nu}= \frac{|{\cal A}(\Psi \to \nu \bar{\nu}) |}
{|{\cal A}(\Psi \to e^+ e^- )| } =  \left( {M_\Psi \over M_Z} \right)^2
\frac{3 - 8\sin^2\theta_w }{16 {\sqrt 2} \cos^2\theta_w \sin^2\theta_w }
\simeq 3.3 \times 10^{-4}~.
\end{equation}
Then, using the theoretical result
$B^{s.d.}(B \to X_s \nu \bar{\nu})=(4.0\pm 1.0)\times 10^{-5}$
\cite{BB2} and the experimental values of $B(B\to X_s \Psi)$ and 
$B(\Psi \to e^+e^-)$ \cite{PDG}, we find
\begin{equation}
\frac{ B(B\to X_s \Psi(\nu \bar{\nu})) }{
B^{s.d.}(B\to X_s \nu \bar{\nu}) } = 3 R_{\nu}^2 \frac{ B(B\to X_s \Psi) 
B(\Psi \to e^+e^-) }{B^{s.d.}(B\to X_s \nu \bar{\nu}) } \simeq 
5 \times 10^{-6}~.
\label{Bnnfinal}
\end{equation}
\\
Unfortunately we cannot reliably estimate the interference terms 
between short-- and long--distance contributions, since we have no 
useful information about the strong phases in $B\to  X_s \Psi$.
However, the result (\ref{Bnnfinal}) indicates that the 
long--distance corrections to the amplitude are at the level of 
$10^{-3}$ at most. This is by far enough to conclude 
that $B \to X_s \nu \bar{\nu}$ is an extraordinarily clean channel 
\cite{BB2} and, 
therefore, a very interesting probe of the Standard Model and its
extensions \cite{GLN}.

\section{Summary}

In this paper we have calculated long--distance corrections of
${\cal O}(\Lambda^2_{QCD}/m^2_c)$ to the dilepton invariant mass
($q^2$) spectrum and to the forward--backward asymmetry in 
$B\to X_se^+e^-$. For low 
$q^2/m^2_b\;\raisebox{-.4ex}{\rlap{$\sim$}} \raisebox{.4ex}{$<$}\; 0.1$
these corrections can be expected to give a fairly accurate 
description of long--distance effects due to intermediate 
(far off--shell) $c\bar c$ pairs. Contributions from 
higher--dimensional operators, formally suppressed by
powers of $m_b \Lambda_{QCD} /m^2_c$ relative to the  
$\Lambda^2_{QCD}/m^2_c$ effect, involve unknown hadronic matrix 
elements and have been neglected. 
Despite the fact that $m_b \Lambda_{QCD} /m^2_c\approx 0.6$, 
this is justified for small $q^2$ due to small numerical 
coefficients accompanying these additional contributions.
They could be more important for larger $q^2$. Close to the resonance
region $q^2/m^2_b\sim 4m^2_c/m^2_b\sim 0.34$ the assumptions on which
the calculation is based clearly break down. On the other hand, simple
order--of--magnitude estimates suggest that even for $q^2/m^2_b\sim 0.26$
the calculable leading ${\cal O}(\Lambda^2_{QCD}/m^2_c)$ effect should
give a reasonable account of the approximate size of the corrections
within, say, a factor of two. Above the resonance region 
$q^2/m^2_b\;\raisebox{-.4ex}{\rlap{$\sim$}} \raisebox{.4ex}{$>$}\; 0.7$
the calculated effect is again reliable.
In spite of the various uncertainties we believe that the $1/m_c$ 
expansion approach is still preferable to model--dependent estimates 
relying on off--shell $c\bar c$ resonances.

Numerically the corrections we find are small (at the one to two percent
level) in the phenomenologically interesting region of $q^2$ away
from the resonances. For comparison one may recall
that uncertainties from QCD perturbation theory are at the level of
$\pm 6\%$ at next--to--leading order.
This indicates that long--distance corrections
related to intermediate $c\bar c$ should not be a serious problem in the
analysis of $B\to X_se^+e^-$ decay, unless the neglected effects of
higher order in $\Lambda^2_{QCD}/m^2_c$ or $m_b\Lambda_{QCD}/m^2_c$
turned out to be substantially larger than anticipated.

We have pointed out that the ${\cal O}(\Lambda^2_{QCD}/m^2_c)$ effect
is calculable in both magnitude and sign, which is to be contrasted
with the situation in hadronic model calculations.
The ${\cal O}(\Lambda^2_{QCD}/m^2_c)$ correction to $B(B\to X_s\gamma)$,
considered in previous work, can be obtained as a special application
of our analysis and is found to be $\approx+3\%$.

The decays $B\to X_s\gamma$ and $B\to X_se^+e^-$ may be contrasted with
the mode $B\to X_s\nu\bar\nu$, where the corresponding nonperturbative
effects due to intermediate $c\bar c$ are still much stronger
suppressed, down to a level of $\sim 10^{-3}$ in the amplitude.

Long--distance effects are in general a notoriously difficult problem.
However,
 they have to be controlled sufficiently well to address fundamental
questions in flavor physics. The existence of a realistic limit in
which such effects can in fact be computed from first principles is
therefore of considerable
theoretical interest in its own right, as well as of
interest for phenomenology. It is particularly gratifying that for the
important processes $B\to X_s\gamma$ and $B\to X_se^+e^-$ a calculation 
of this type is possible and a class of potentially large long--distance
corrections is indeed found to be rather well under control.

{\bf Note added:} 
After this paper was finished, we have received a preprint \cite{CRS}
in which the ${\cal O}(\Lambda^2_{QCD}/m^2_c)$ correction 
to the differential decay rate of $B \rightarrow X_s e^+e^-$
has been calculated. We disagree with their result in sign. 

\section*{Acknowledgements}
We are grateful to P. Kim for illuminating discussions concerning
experimental issues of $B\to X_s\gamma$ decay and to
Z. Ligeti, M. Voloshin, M. Wise and D. Wyler for 
informative discussions about their work. We also thank 
C. Greub, Y. Grossman, H. Quinn and T. Rizzo for discussions.


\begin{thebibliography}{99}
\bibitem{GHW}
C. Greub, T. Hurth and D. Wyler, Phys. Rev. {\bf D54}, 3350 (1996).
\bibitem{CMM}
K.G. Chetyrkin, M. Misiak and M. M\"{u}nz, ZU-TH-24-96, hep-ph/9612313.
\bibitem{MIS}
M. Misiak, Nucl. Phys. {\bf B393}, 23 (1993); erratum ibid.
{\bf B439}, 461 (1995).
\bibitem{BM}
A.J. Buras and M. M\"{u}nz, Phys. Rev. {\bf D52}, 186 (1995).
\bibitem{BB2}
G. Buchalla and A.J. Buras, Nucl. Phys. {\bf B400}, 225 (1993).
\bibitem{FLS}
A.F. Falk, M. Luke and M.J. Savage, Phys. Rev. {\bf D49}, 3367 (1994).
\bibitem{AHHM}
A. Ali, G. Hiller, L.T. Handoko and T. Morozumi, 
Phys. Rev. {\bf D55}, 4105 (1997).
\bibitem{BBSUV}
I.I. Bigi {\it et al.}, in {\it B-Decays} (2nd edition),
ed. S.L. Stone, World Scientific, Singapore (1994), p. 132.
\bibitem{GLN}
Y. Grossman, Z. Ligeti and E. Nardi, Nucl. Phys. {\bf B465}, 369 (1996);
erratum ibid. {\bf B480}, 753 (1996).
\bibitem{HW}
J.L. Hewett and J.D. Wells, Phys. Rev. {\bf D55}, 5549 (1997).
\bibitem{DHT}
N.G. Deshpande, X.-G. He and J. Trampeti\'{c},
Phys. Lett. {\bf B367}, 362 (1996); \\
G. Eilam, A. Ioannissian, R.R. Mendel and P. Singer, 
Phys. Rev. {\bf D53}, 3629 (1996).
\bibitem{SOA}
J.M. Soares, Phys. Rev. {\bf D53}, 241 (1996).
\bibitem{AMM}
A. Ali, T. Mannel and T. Morozumi, 
Phys. Lett. {\bf B273}, 505 (1991).
\bibitem{AH}
M.R. Ahmady, Phys. Rev. {\bf D53}, 2843 (1996);
C.-D. L\"{u} and D.-X. Zhang, Phys. Lett. {\bf B397}, 279 (1997).
\bibitem{VOL}
M.B. Voloshin, Phys. Lett. {\bf B397}, 275 (1997).
\bibitem{KRSW}
A. Khodjamirian, R. R\"uckl, G. Stoll and D. Wyler,
WUE-ITP-97-001, hep-ph/9702318.
\bibitem{LRW}
Z. Ligeti, L. Randall and M.B. Wise, CALT-68-2097, hep-ph/9702322.
\bibitem{GMNP}
A.K. Grant, A.G. Morgan, S. Nussinov and R.D. Peccei,
UCLA/97/TEP/5, hep-ph/9702380.
\bibitem{NEU}
M. Neubert, Phys. Rev. {\bf D49}, 4623 (1994); J. Chay and S.-J. Rey,
Z. Phys. {\bf C68} (1995) 425. 
\bibitem{AG}
A. Ali and C. Greub, Phys. Lett. {\bf B361}, 146 (1995).
\bibitem{CLEO} M.S. Alam {\it et al.} (CLEO Collaboration),
Phys. Rev. Lett. {\bf 74}, 2885 (1995). 
\bibitem{DEIN} 
G. D'Ambrosio, G. Ecker, G. Isidori  and H. Neufeld, 
Phys. Lett. {\bf B380}, 165 (1996); 
G. D'Ambrosio and G. Isidori, LNF-96/036(P), hep-ph/9611284.
\bibitem{AGbrem}
A. Ali and C. Greub, Z. Phys. {\bf C49}, 431 (1991);
{\it ibid.} {\bf C60}, 433 (1993); 
N. Pott, Phys. Rev. {\bf D54}, 938 (1996).
\bibitem{CFRS}
M. Ciuchini, E. Franco, L. Reina and L. Silvestrini,
Nucl. Phys. {\bf B421}, 41 (1994).
\bibitem{BBL}
G. Buchalla, A.J. Buras and M.E. Lautenbacher,
Rev. Mod. Phys. {\bf 68}, 1125 (1996).
\bibitem{FN} 
A.F. Falk and M. Neubert, Phys. Rev. {\bf D47}
(1993) 2965.
\bibitem{JL}
M. Jamin and M.E. Lautenbacher, Comput. Phys. Commun. {\bf 74}, 265 (1993).
\bibitem{LW} 
Z. Ligeti and M.B. Wise, Phys. Rev. {\bf D53}, 4937 (1995).
\bibitem{KS}
F. Kr\"{u}ger and L.M. Sehgal, Phys. Lett. {\bf B380}, 199 (1996).
\bibitem{PDG} 
Review of Particle Properties, Phys. Rev. {\bf D54}, 1 (1996).
\bibitem{CRS} J.-W. Chen, G. Rupak and M.J. Savage, DOE/ER/41014-10-N97,
hep-ph/9705219.

\end{thebibliography}
\end{document}